\documentclass{osa-article}

\journal{oe}

\articletype{Research Article}

\begin{document}

\title{Extraction and identification of noise patterns for ultracold atoms in an optical lattice}

\author{Shuyang Cao,\authormark{1} Pengju Tang,\authormark{1} Xinxin Guo,\authormark{1} Xuzong Chen,\authormark{1} Wei Zhang,\authormark{2,*} and Xiaoji Zhou\authormark{1,$\dagger$}}

\address{\authormark{1}School of Electronics Engineering and Computer Science, Peking University, Beijing 100871, China\\
\authormark{2}Department of Physics, Renmin University of China, Beijing 100872, China}

\email{\authormark{*}wzhangl@ruc.edu.cn\\
\authormark{$\dagger$}xjzhou@pku.edu.cn} 



\begin{abstract}
To extract useful information about quantum effects in cold atom experiments, one central task is to identify the intrinsic quantum fluctuation from extrinsic system noises of various kinds. As a data processing method, principal component analysis can decompose fluctuations in experimental data into eigen modes, and give a chance to separate noises originated from different physical sources. In this paper, we demonstrate for Bose-Einstein condensates in one-dimensional optical lattices that the principal component analysis can be applied to time-of-flight images to successfully separate and identify noises from different origins of leading contribution, and can help to reduce or even eliminate noises via corresponding data processing procedures. The attribution of noise modes to their physical origins is also confirmed by numerical analysis within a mean-field theory. 
\end{abstract}

\section{\label{introduction}Introduction}
Cold atom systems provide a unique opportunity to obtain various information with high precision from an interacting many-body system, which can help us gaining physical understanding of strongly correlated systems. For instance, analysis of spatial noise correlations in the time-of-flight (TOF) images can reveal density or spin correlations for atoms loaded in optical lattices (OLs)~\cite{folling2005spatial, Altman}, or pairing correlation in a Fermi superfluid~\cite{Altman, Greiner}. Measurements of in-situ density fluctuations have revealed the Pauli blocking effect~\cite{Shin}, provided information of spin or density susceptibility of strong interacting gases~\cite{Sanner}, and verified the scaling law of a critical state in two-dimensional Bose gas~\cite{Gemelke}. 

To reveal quantum correlation and fluctuation effects in an ultracold atomic gas, a central task is to separate the extrinsic noises due to the imperfect experimental setup and state preparation from the intrinsic sources of quantum or thermal fluctuations. These noises are often coupled, warped by nonlinear effects, and buried in massive pixels, which make the task even harder. Principal component analysis (PCA) provides a great approach for solving this problem~\cite{LEVINE, PCANJP1, PCANJP2, PCAPRA, Jolliffe}. Nowadays this method is being widely used in computer science to help reducing data dimensions and find internal structure of massive high dimensional data. For a similar purpose, we use PCA to analyze time-of-flight (TOF) images of Bose-Einstein condensates (BECs) in one-dimensional (1D) OLs., where data dimensions are as many as the number of image pixels but the noise origins are much fewer. The application of PCA on the raw TOF data suggests that the leading noise sources in our experiment are fluctuations of atom number and spatial position. By preprocessing the raw data with normalization and adaptive region extraction methods, we can significantly reduce or even eliminate these two noises. As a result, PCA of the preprocessed data reveals more subtle structure of noises. We attribute the few dominant noise components with their corresponding physical origins, and compare experimental results with numerical simulations using physical parameters determined by experiments. 

The core of PCA is to represent variations approximately using a minimal group of orthogonal vectors called {\it principal components} (PCs) while preserving most of information, by which we can retain main features of variations without being distracted by other less essential factors. As PCA is applied to experimental data, PCs acquire their physical meanings apart from their original concepts in mathematics. In our system, experimental data are time-of-flight images and PCs are effectively eigen modes of fluctuations in our experiments. Thus, the total fluctuation is a linear combination of these eigen modes, and the result of a specific TOF image denoted by $A_i$ can be represented by the average over images plus its fluctuation, namely $A_{i} = \bar{A} + \sum\varepsilon_{ij}P_{j}$. Here, $P_j$ denotes different eigen modes of fluctuation, the absolute value of the coefficient $\varepsilon_{ij}$ describes how much $P_{j}$ contributes to the \textit{i}-th measurement $A_{i}$, and its sign indicates in which way $P_{j}$ influences $A_{i}$.

Because of the linearity of PCA, the eigen modes associated with different PCs have a one-to-one correspondence to different sources of fluctuations in a linear system. For a nonlinear system, such as an interacting many-body quantum system, where variations are warped by nonlinear interaction effects, the eigen modes of PCA are in general nonlinear combinations of various noises. However, as we will demonstrated below, the nonlinearity in the present system turns out to be sufficiently small, such that different noises can be decomposed efficiently using PCA. 

The remainder of this manuscript is organized as follows. In Sec.~\ref{expe_setup}, we briefly introduce the experimental setup. The protocols of the PCA method is explained in Sec.~\ref{analysis_method}, and then implemented for BECs both in the absence and presence of OL in Secs.~\ref{sec:bec_analysis} and \ref{sec:ol_analysis}, respectively. By comparing with numerical simulation, we identify the physical origins of up to five dominant PCs in the noise of TOF images. Finally, we discuss some remarks of the PCA method and summarize in Sec.~\ref{pca_discussion}.

\section{\label{expe_setup}Experimental setup}
\begin{figure}[tbp!]
\centering\includegraphics[width=11cm]{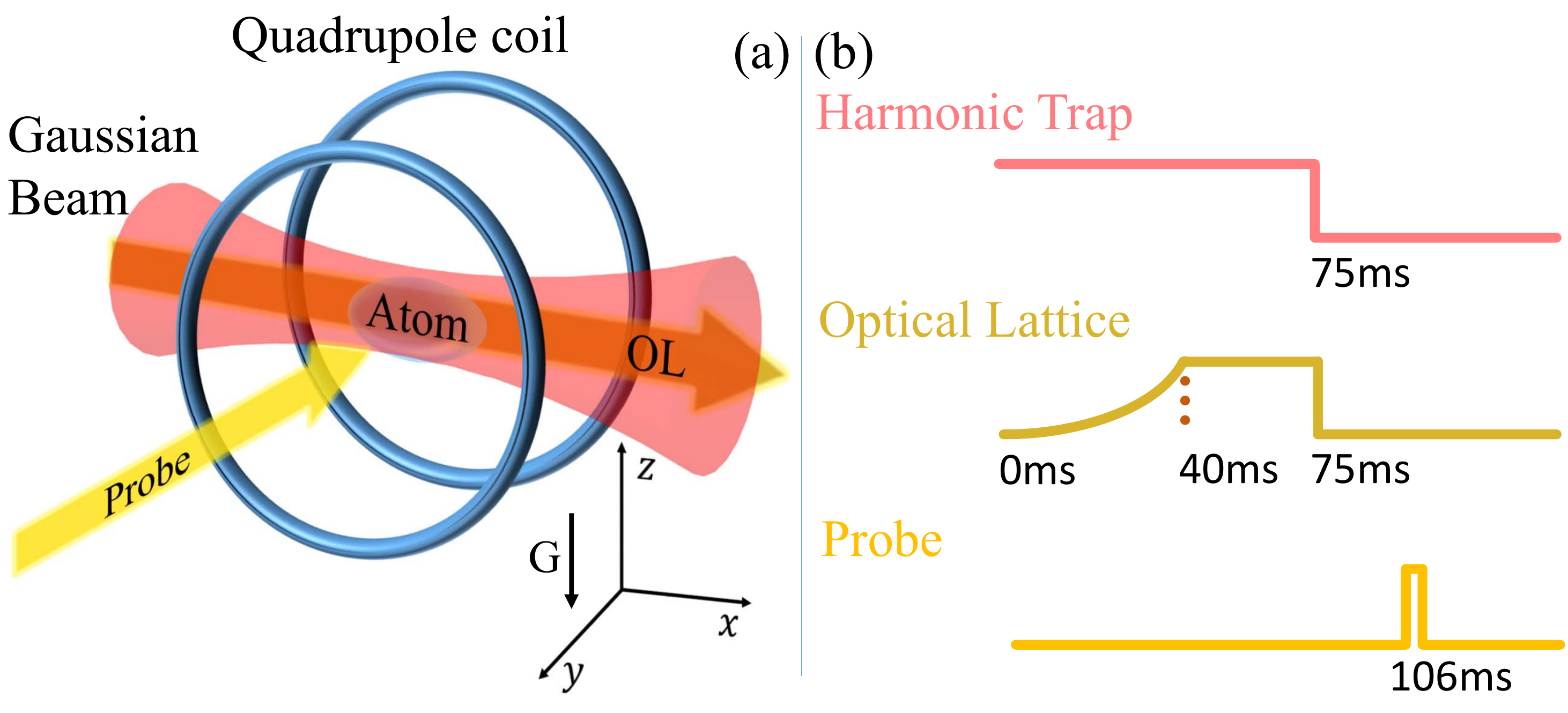}
\caption{\label{fig:sys_struc} (a) Experimental setup. The harmonic trap is composed of potentials produced by the quadruple coil and the Gaussian beam. The direction of the optical lattice (OL) is roughly the same as the Gaussian beam and they overlap at the position of the atomic cloud. (b) Trap time sequence. BEC is prepared in a harmonic trap first. Then OL starts to ramp up adiabatically at $t = 0$ ms and reaches the configured depth in 40 ms. After a holding time of 35 ms, both the harmonic trap and the OL are turned off at the same time. Absorption images are taken in the $x$-$z$ plane upon 31ms of free expansion.}
\end{figure}

The system we used here is similar to the one in our previous experiments \cite{PhysRevA.94.043607, PhysRevA.94.063603, PhysRevA.94.033624}, which is a hybrid trap composed of a quadrupole magnetic trap and an optical dipole trap, as shown in Fig.~\ref{fig:sys_struc}. Our BEC setup is as follows. A BEC of about $N_{0} = 2\times 10^{5}$ ${}^{87}$Rb atoms in the $|F = 2, m_{F} = 2\rangle$ state is first prepared in the trap with frequencies $\omega_{x} = 2\pi \times$ 28Hz, $\omega_{y} = 2\pi \times$ 50 Hz and $\omega_{z} = 2\pi \times60$ Hz.
Within 40 ms the BEC is adiabatically loaded into a 1D OL along the $x$-direction. The lattice wavelength is 852 nm, and the height can be tuned within a range from $6{E}_{R}$ to $21{E}_{R}$, where ${E}_{R} = \hbar^{2}k_{L}^{2}/2m$ is the photon recoil energy. After 35 ms we turn off the harmonic trap and the OL simultaneously to release the BEC. The absorption images are taken in the $x$--$z$ plane upon 31ms of free expansion with the size of each CCD pixel $6.8\mu$m$\times6.8\mu$m. Here, we use strong saturated near-resonance imaging laser to obtain the density distribution of atomic gas of high density, and calibrate the the imaging system to validate the TOF measurement~\cite{saturatedOD}. Since absorption imaging is destructive, atoms have to be prepared repeatedly, resulting in variations from shot to shot inevitably. In our experiments, we took about 40 images at each lattice depth. We also studied 100 TOF images of BEC without OL as a preliminary experiment. For each raw TOF image, we implement an optimized fringe removal algorithm (OFRA) to eliminate the background interference of imaging light~\cite{OFRA}. As a result, the residue fringes are much weaker in magnitude and can be easily discriminated from signals for the few leading principal components.

\section{\label{analysis_method}Methods of data analysis}

\subsection{\label{subsec:pca_steps}Protocol of PCA}

First, we need to extract an $h\times w$ region of interest from the raw TOF images, and then apply PCA to those regions. The region should be as small as possible to reduce noises, but fully cover the area where atoms reside. Details of the method are described below and illustrated in Fig.~\ref{fig:schema}:

\begin{figure*}[ht!]
\centering\includegraphics[width=11cm]{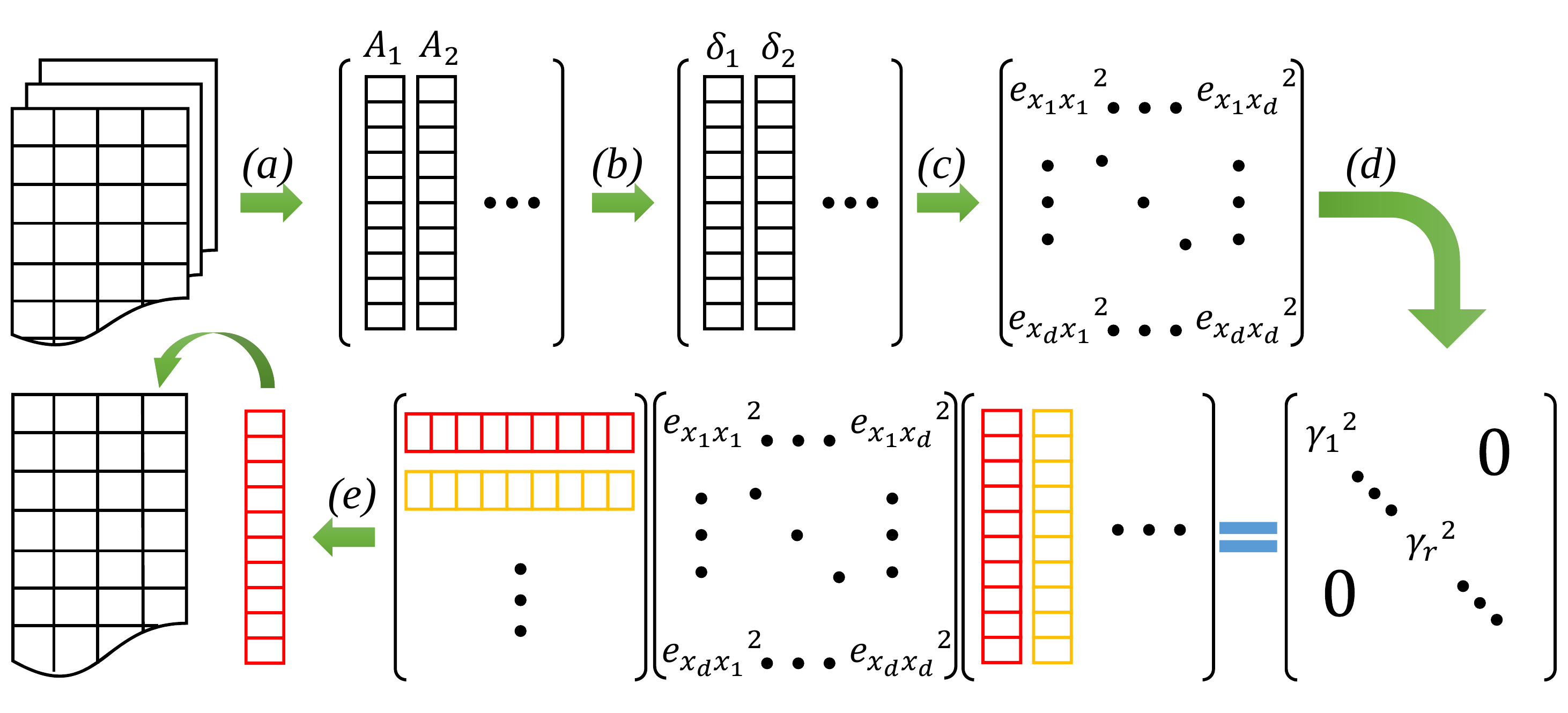}
\caption{\label{schema}PCA protocols. (a) Transform two-dimensional regions of interest into column vectors $A_i$ and stack them together. (b) Subtract the mean vector from $A_i$ and keep fluctuations $\delta_i=A_i-\bar{A}$. (c) Construct covariance matrix $S = \frac{1}{n}X\cdot X^{T}$. (d) Decompose covariance matrix so that ${{V}^{-1}{S}{V} = {D}}$ where D is a diagonal matrix. (e) Transform eigen vectors of interest back to two-dimensional regions to reconstruct images.}
\label{fig:schema}
\end{figure*}

\textbf{\textit{Vectorize image}} For the $i$-th image, retrieve the region of interest and transform it into a $d$-dimensional vector $A_i$, where $d = h\times w$.

\textbf{\textit{Decompose $A_i$}} Denote $A_i = \bar{A} + \delta_{i}$, where $\bar{A}= \frac{1}{n}\sum_{i=0}^{n} A_i$ is the average of all $n$ vectors, and $\delta_{i} = A_{i} - \bar{A}$ is the fluctuation.

\textbf{\textit{Construct covariance matrix}} Stack $\delta_{i}$ together to form a matrix $X = \left[\delta_{1},\delta_{2},\cdots,\delta_{n}\right]$. Then the covariance matrix $S$ is obtained by $S = \frac{1}{n}X\cdot X^{T}$.

\textbf{\textit{Decompose covariance matrix}} Compute the matrix $V$ of eigenvectors which diagonalizes the covariance matrix with ${{V}^{-1}{S}{V} = {D}}$.

\textbf{\textit{Reconstruct images}} If needed, reshape those $d\times 1$ eigenvectors of interest back to $h\times w$ matrices to reconstruct feature images.

We use the Scree graph method \cite{Jolliffe} to determine the number of PCs to be retained. Specifically, we plot the eigenvalues in descending order and determine the turning point from which the curve flattens. The eigenvalues above the turning point are of significance and about to be retained.

\subsection{\label{subsec:processing_methods}Preprocessing Method}

To identify the physical origins of the PCs, we also preprocess the raw TOF data to 
eliminate the fluctuations of atom number and spatial position, and compare the new outcome of PCA with the 
original ones. The preprocessing methods are listed as follows.

\paragraph{Normalization}
Normalization is designed to reduce atom number fluctuation in the region of interest. In TOF images, the atom number is 
determined as
\begin{eqnarray}
N=-\sum{\ln{\left(\frac{I-I_{bg}}{I_0-I_{bg}}\right)}\frac{s}{\sigma}},
\end{eqnarray} 
where $s$ is the CCD pixel size, $\sigma$ is the absorption cross section, 
$I$ is the CCD pixel value when probing atoms, $I_0$ is the pixel value when there is probe light but no atoms, 
and $I_{bg}$ is the background when there is no light. 

As weakly interacting BECs in our experiments can be described by a macroscopic wave function $\Psi{\left(\textbf{r}\right)}=\sqrt{N}\phi{\left(\textbf{r}\right)}$~\cite{macroscopic}, our interest is the density distribution of the normalized wave function ${\left|\phi{\left(\textbf{r}\right)}\right|}^2$ instead of the prefactor $N$.  So it is safe to normalize the TOF data so that
the pixel values in a region of interest are summed up to unity. By doing so, densities in different TOF images fall into the same range and become comparable. 

To normalize the raw TOF data, we first eliminate the bias. In principle, the pixel values at positions where no atom is present should be zero. However, there may exist a finite signal as a bias in real experiments. One method to eliminate this noise is by simply taking the bias as the minimal value of pixels. A slightly more complicated but more robust method that is used in our experiment is taking the mean value of pixels where there is no atom as a uniform background noise, and then subtracting it from all pixels. After the elimination of bias, we can normalize the pixel values using
\begin{eqnarray}
{\tilde v}_{ij}=\frac{v_{ij}}{\sum_{i}\sum_{j}v_{ij}},
\end{eqnarray}
where $v_{ij}$ denotes the raw intensity of pixel labeled by coordinate indices $i$ and $j$.

\paragraph{Adaptive region extraction}

Another fluctuation of TOF images is the shift of the cloud position, which may be induced by experimental misalignments of trapping potential, optical lattice potential, or imaging camera. Adaptive region extraction is designed to select a region whose center is also the center of density distribution. We first set a criterion to determine the center of density distribution within an extracted region, then use the center as a new region center to extract a new region. We iterate this procedure until the region to be extracted becomes stable. Note that in general the  coordinates of the cloud center are not integers. While simply rounding them to integers may cause artificial anisotropy in our images, we use interpolation to estimate the pixel values with non-integer coordinates.

For the choice of criterion, a simple method is to set the pixel with maximal value as the center of density distribution. This procedure works well in most cases provided that the CCD pixel noises are small enough. In the following discussion, however, we use a slightly generalized criterion which determines the center by a weighted mean of all pixels in the region, where the weight is chosen to be the pixel values.

\section{\label{sec:bec_analysis} PCA Analysis in the Absence of OL}

\begin{figure}[ht!]
\centering\includegraphics[width=11cm]{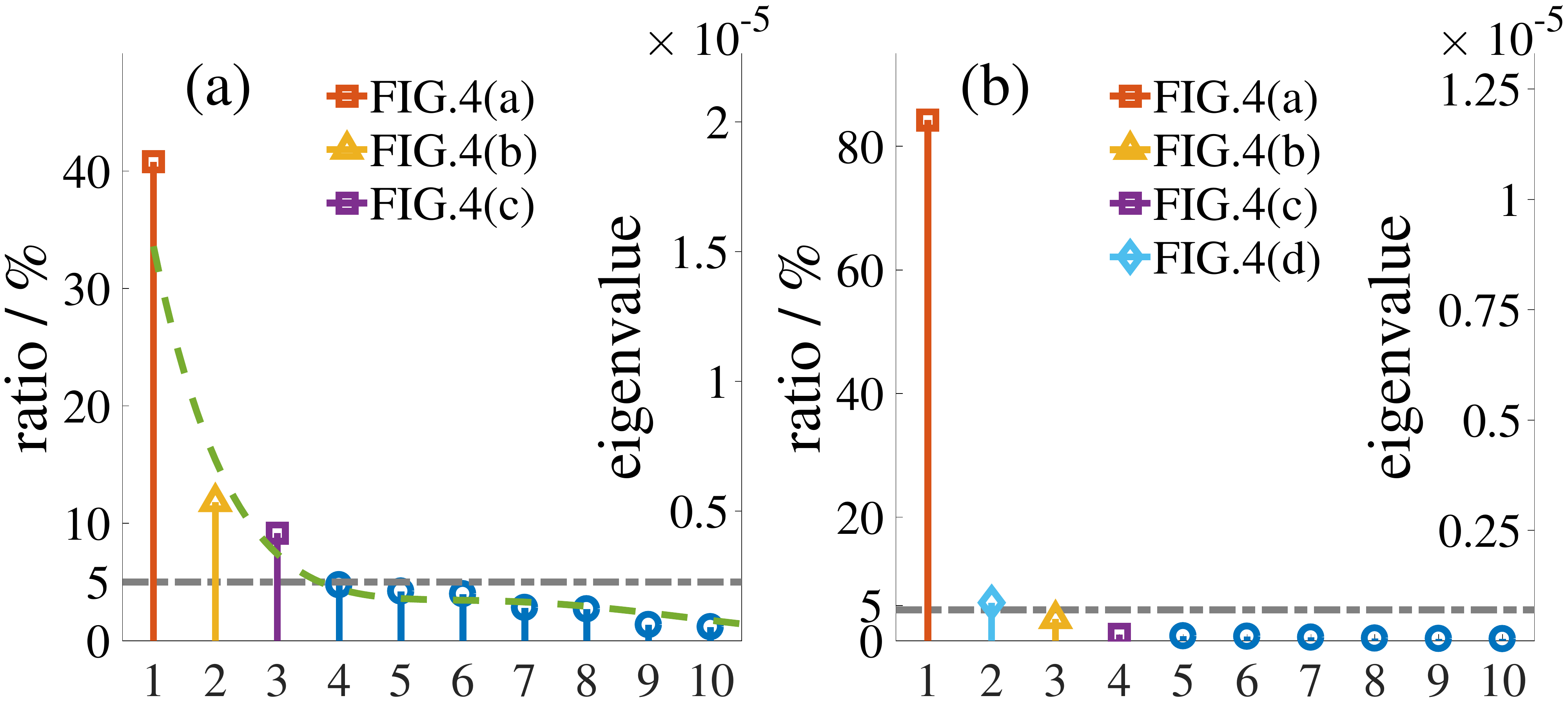}
\caption{The magnitude (right axis) and percentage ratio (left axis) of the eigenvalues associated to the first 10 PCs before (a) and after(b) preprocessing. The ratio is defined as the percentage of corresponding eigenvalue out of the summation of all eigenvalues. The green dashed line is a smoothed line that connects the first ten points to help find the turning point. The gray dashed line indicates the threshold for distinguishing important PCs. In our experiments, the threshold is 5\%. PCs of interest are highlighted with different colors. Note that the first three PCs are significantly reduced in magnitude by preprocessing. Meanwhile, a new PC depicted by Fig.~\ref{fig:bec_pca_img}(d) appears after preprocessing as other leading noises sources are strongly suppressed.}
\label{fig:bec_pca_contrast_new_img}
\end{figure}

\begin{figure}[ht!]
\centering\includegraphics[width=11cm]{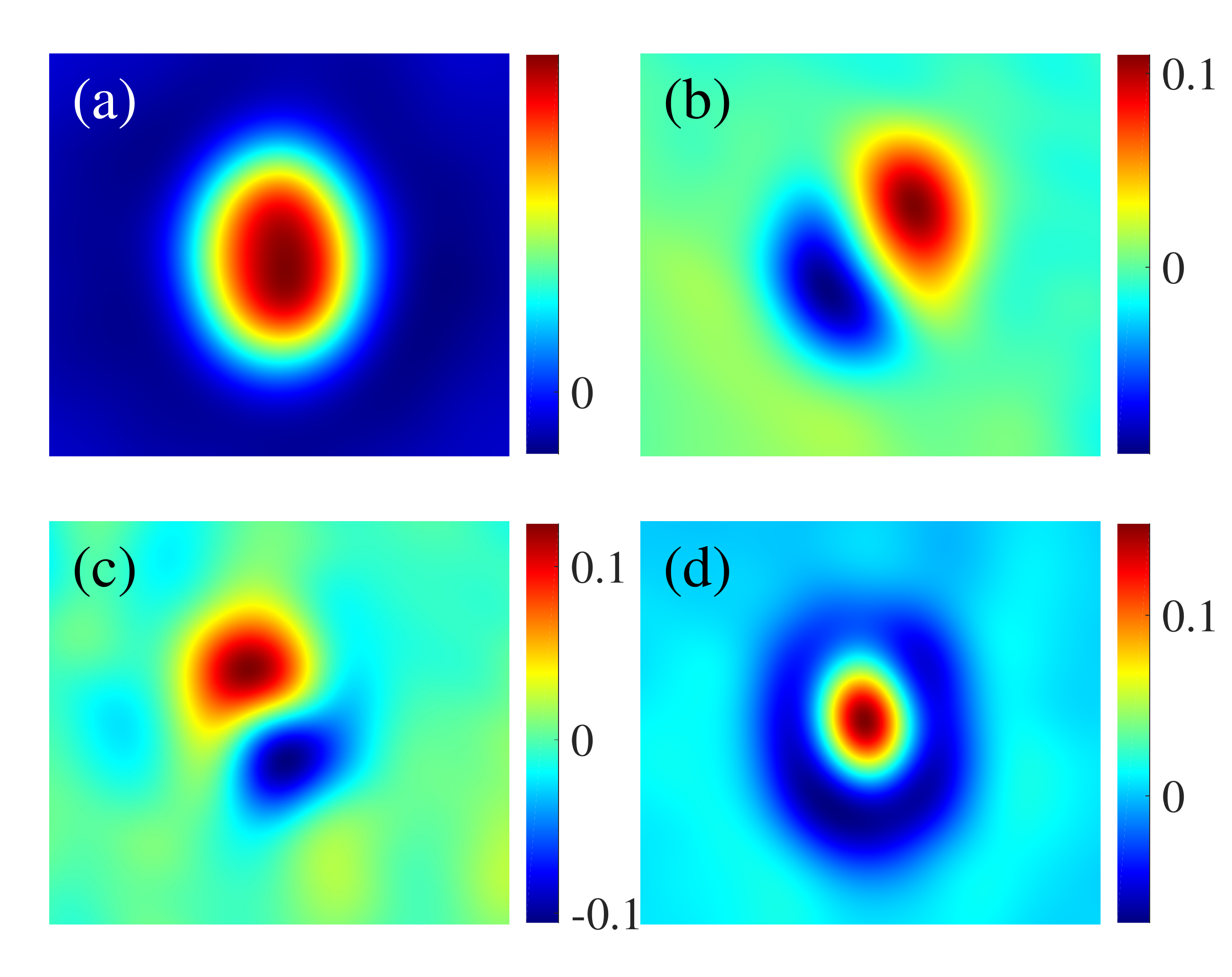}
\caption{Reconstructed feature images. (a) represents the fluctuation of atom number. (b) and (c) correspond to the spatial fluctuations on $x$- and $z$-directions, respectively. (d) The new PC after preprocessing is characterized by a peak wrapped by a dip ring.}
\label{fig:bec_pca_img}
\end{figure}

As a preliminary experiment, we first analyze TOF absorption images of a 31ms free-expanding BEC released from the harmonic trap, whose fluctuation modes should be relatively simpler. Fig.~\ref{fig:bec_pca_contrast_new_img}(a) shows percentages of the total variance associated with the first 10 PCs. The green dashed line is a smoothed line that connects these ten points, from which we can easily tell the turning point resides between the third and fourth PCs and the critical value for retaining PCs is around $5\%$. So we reconstruct the feature images corresponding to the first three primary PCs, which are shown in Figs.~\ref{fig:bec_pca_img}(a)-\ref{fig:bec_pca_img}(c). Figure~\ref{fig:bec_pca_img}(a) is similar to the original absorbing image, which corresponds to atom number fluctuations. Figures~\ref{fig:bec_pca_img}(b) and \ref{fig:bec_pca_img}(c), whose percentages are of the same order of magnitude, reflect the position uncertainty of the BEC along the $x$- and $z$-directions, respectively. Their origin should be some mechanical effects such as a shift of the magnetic trap position or drift of the CCD camera, both of which can cause a position deviation of BEC in TOF images.

To validate our attribution, we preprocess data using methods introduced in Sec.~\ref{subsec:processing_methods} to 
eliminate the number and position fluctuations, which respectively correspond to the feature images
of PCs shown in Fig.~\ref{fig:bec_pca_img}(a) and Figs.~\ref{fig:bec_pca_img}(b-c). We then apply PCA to the 
resulting data, and plot the percentages and eigenvalues of the first 10 leading PCs in Fig.~\ref{fig:bec_pca_contrast_new_img}(b). 
As compared to the outcome without preprocessing, there are only two PCs left above the critical line (gray line in Fig.~\ref{fig:bec_pca_contrast_new_img}(b)), with a combined contribution of $>90\%$. We reconstruct feature images of the first four PCs, and find that PCs whose patterns are similar to Figs.~\ref{fig:bec_pca_img}(a-c) now take the first, third and fourth places. The eigenvalue of the PC depicted in Fig.~\ref{fig:bec_pca_img}(a) decreases by $36\%$, from $1.846\times{10}^{-5}$ to $1.180\times{10}^{-5}$, even if its ratio becomes larger because other noises are suppressed stronger. The eigenvalues and ratios of PCs depicted in Figs.~\ref{fig:bec_pca_img}(b) and \ref{fig:bec_pca_img}(c) are below the critical line at present, ready to be ignored in our analysis. A new PC takes the second place after preprocessing, corresponding to the variation of the transversal radius of the cloud as depicted in Fig.~\ref{fig:bec_pca_img}(d), which may be attributed to the breathing mode of BEC. 

From this result, we can conclude that the atom number fluctuations are effectively reduced and the position fluctuations are nearly eliminated by our preprocessing. We stress that if one employs a simpler but coarse version of adaptive region extraction where the pixel coordinates is always rounded to integers, although the PCs associated with position fluctuations can still be significantly reduced in magnitude, they remain to be leading PCs as the truncation errors are relevant if the size of our CCD pixels are larger than the real spatial shifts.

\section{\label{sec:ol_analysis}PCA Analysis of BEC in OL}

We now apply PCA to analyze TOF images of BECs in an OL. In Sec.~\ref{sec:ol_pca_results}, 
we present the first five leading PCs and their corresponding feature images for a typical optical lattice depth 
of $15E_R$. The experimental results are in good quantitative agreement with numerical simulation as 
discussed in Sec.~\ref{subsec:simu_pca_ol}. Finally, we discuss in Sec.~\ref{subsec:depth_trend} 
the variation of the leading PCs with optical lattice depth.

\subsection{\label{sec:ol_pca_results}PCA results of BEC in OL}

\begin{figure}[ht!]
\centering\includegraphics[width=11cm]{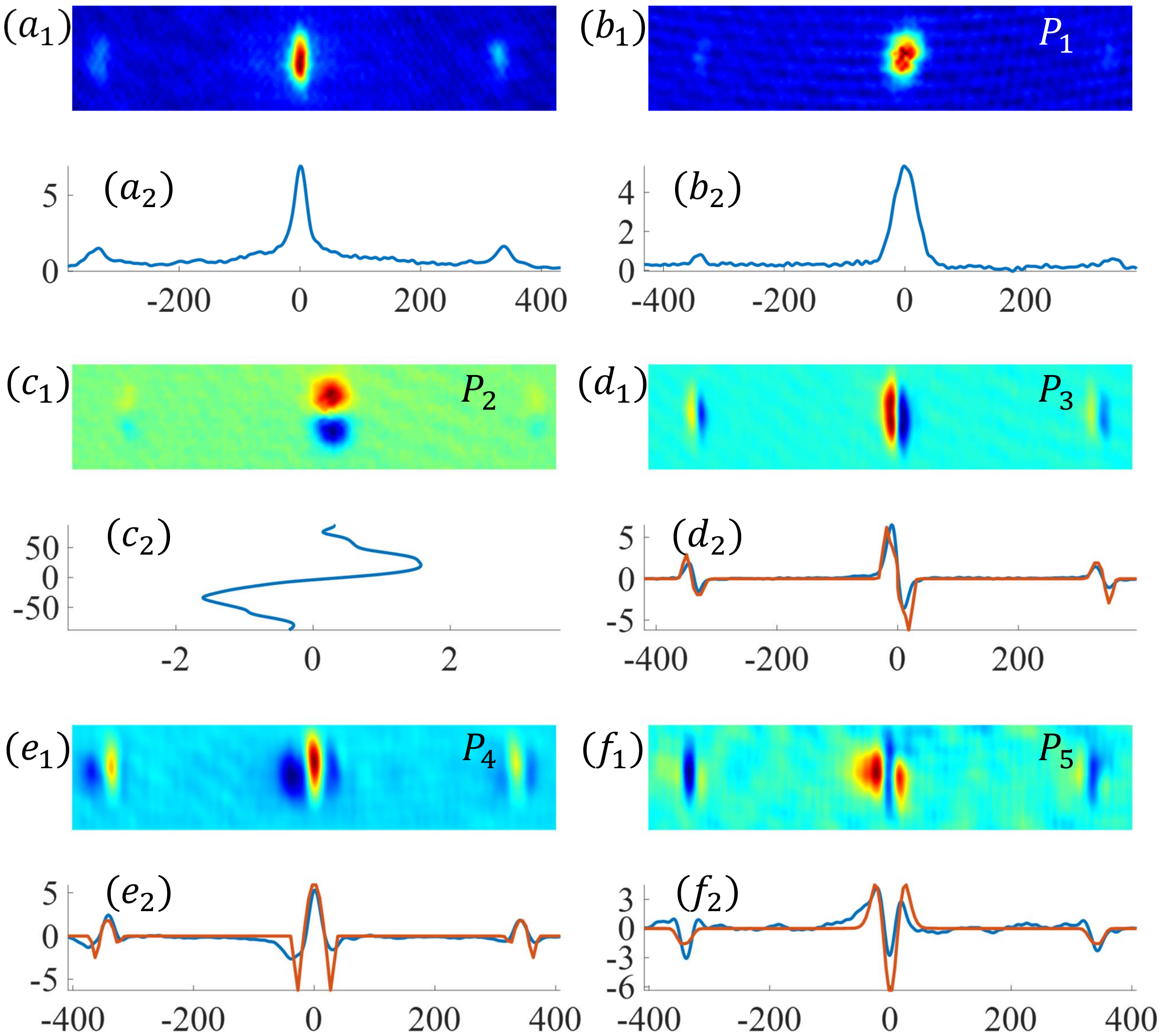}
\caption{PCA results of BEC in OL. ($a$) Raw TOF image. ($b$) Atom number fluctuation. ($c$) and ($d$) Position fluctuations. ($e$) Peak width fluctuation. ($f$) Normal phase fraction fluctuation. We integrate the results of ($a_1$), ($b_1$), ($d_1$), ($e_1$), ($f_1$) vertically and ($c_1$) horizontally, to obtain the columnar integral as depicted by blue lines in the bottom parts of (a-f). Orange lines are simulation results. The unit of horizontal axes in ($a_2$), ($b_2$), ($d_2$), ($e_2$), ($f_2$) and the vertical axis in ($c_2$) are $\mu m$.}
\label{fig:ol_pca_img}
\end{figure}

We analyze TOF images of BECs loaded in an OL with depth of $15E_R$ with the PCA method introduced 
above, and use the Scree graph method to obtain the feature images of the first five leading 
PCs as shown in the top panels of Fig.~\ref{fig:ol_pca_img}(b-f). To see clearly the variation patterns of 
these images, we also integrate over the vertical (horizontal) dimension for Figs.~\ref{fig:ol_pca_img}(b) 
and \ref{fig:ol_pca_img}(d-f) (Fig.~\ref{fig:ol_pca_img}(c)) by summing up pixel values, 
and show the columnar density by solid blue lines in the corresponding bottom panels. 
In the top panel of Fig.~\ref{fig:ol_pca_img}(a), 
we present a typical example of TOF image before preprocessing, while the blue line in the bottom 
shows the three interference peaks clearly.

The first PC, denoted by $P_1$ (Fig.~\ref{fig:ol_pca_img}(b1)), has the same pattern as the atom number fluctuation PC 
discussed in Sec.~\ref{sec:bec_analysis}, except for the symmetric side peaks caused by the presence of OL. 
Indeed, even in the case of high lattice depth where the atoms residing on different lattice sites form 
a local quasi-condensate while the system as a whole does not possess long-range phase coherence, the 
side peaks are still present as a consequence of short-range correlation~\cite{weizhang}. 
If we normalize the data, this PC will disappear or become significantly less important.

The second ($P_2$) and third ($P_3$) PCs as depicted in Figs.~\ref{fig:ol_pca_img}(c1) and \ref{fig:ol_pca_img}(d1) show clear patterns of position fluctuation along the $z$- and $x$-directions, 
respectively. If we extract the region of interest adaptively, these noise modes almost disappear. 
However, unlike the case without OL, if we eliminate the first three PCs with preprocessing method, a PCA on the resulting data gives a leading PC of number fluctuation mode again, indicating that the number fluctuation of BEC in OLs can not be normalized as effectively as in the case without OL.

The feature image of the forth PC $P_4$ as shown in Fig.~\ref{fig:ol_pca_img}(e1) is similar to that of the 
number fluctuation mode $P_1$, but with two negative dips accompanying the interference peaks. 
As we will see in the next subsection, this pattern reflects fluctuations of the width of each peak in the TOF image. 

The fifth PC ($P_5$ as in Fig.~\ref{fig:ol_pca_img}(f1)) is featured by a central dip and two side dips with an overall Gaussian profile. The profile strongly suggests an intimate relation to fluctuations of the normal fluid fraction. The presence of dips can be understood by noticing that within the constraint of atomic number conservation, the increase of thermal atomic number is accompanied by a decrease of condensation fraction, which in turn leads to a reduced visibility of interference pattern. 

Before concluding this subsection, we emphasize that the patterns of $P_4$ and $P_5$ are very small variations which can strongly couple with background noises. It is very difficult to distinguish these noises by conventional analysis on the TOF images. As a comparison, PCA works very effectively to extract these information.

\subsection{\label{subsec:simu_pca_ol}Numerical simulation and comparison with experiments}

To validate our previous attribution of physical origins to different PCs, we perform a numerical simulation for BEC in an OL using time-split spectral algorithm (TSSP) algorithm~\cite{Bao_Ground, Bao_Dynamic}. We first calculate the ground state wave function of the BEC by solving the conventional Gross-Pitaevskii equation (GPE) within the potential generated by the OL and the magneto-optical hybrid trap. The wave function then undergoes a free expansion of 31ms governed by the time-dependent GPE. At the end, a Gaussian envelop is added to the density distribution to simulate the excited fraction, which can not be described properly with the GPE. The reason we use a Gaussian distribution to describe non-condensed particles is because the velocities of these atoms obey a Maxwell distribution after release.

\begin{figure}[ht!]
\centering\includegraphics[width=11cm]{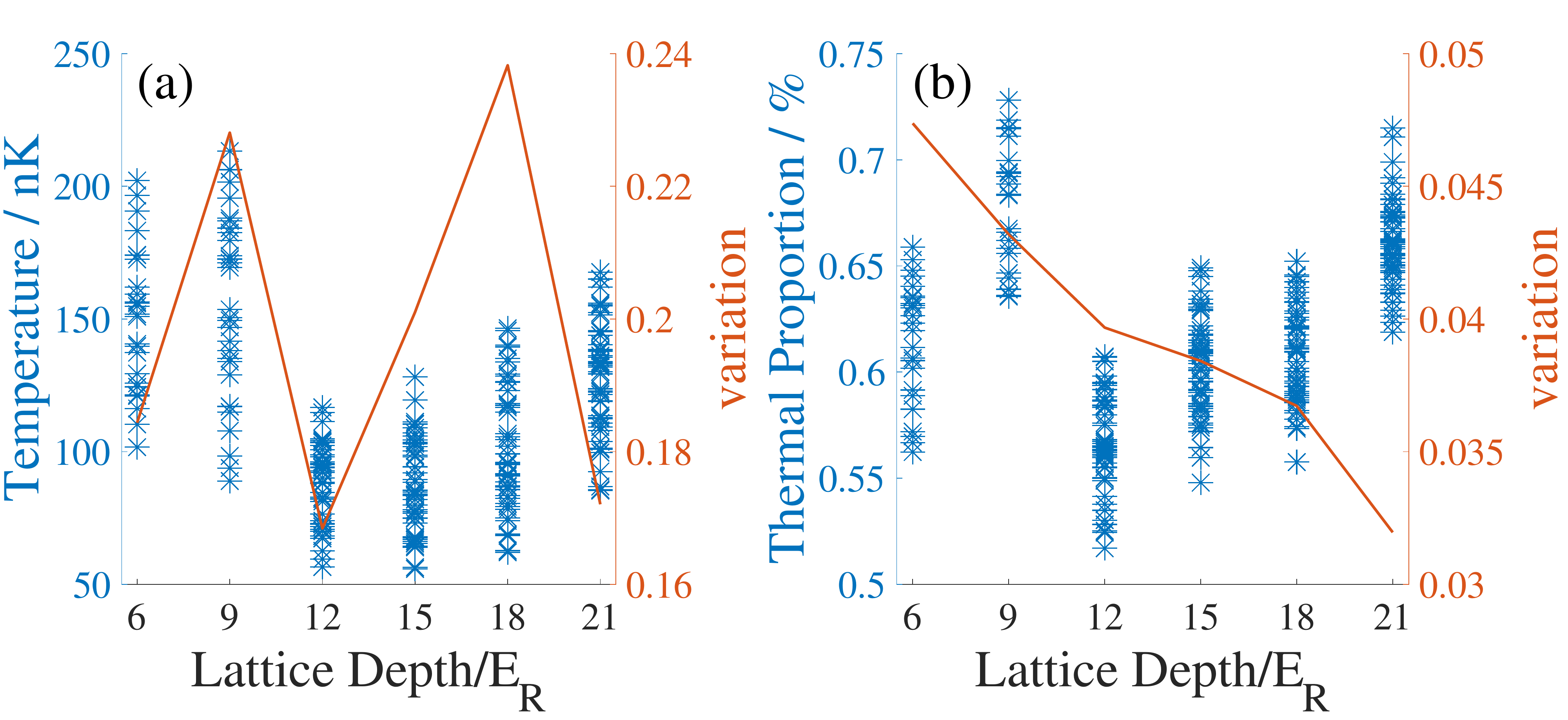}
\caption{Statistical results of experiments. Each asterisk in the figure corresponds to one experimental shot. (a) Statistical results of temperature. (b) Statistical results of thermal atom fraction, which is defined as ${N_{\rm thermal}}/{N_{\rm tot}}$. The orange lines in panel (a) and (b) indicate the variation of the corresponding quantity. The variation is defined as the standard deviation divided by the mean.}
\label{fig:T_and_percent}
\end{figure}

To incorporate the fluctuation effects in our simulation, we extract the thermal fraction and the temperature of each shot by a bimodal fit of the raw TOF images. From Fig.~\ref{fig:T_and_percent}, we find that the fluctuations of temperature $T$ and thermal atom fraction $P_{\rm ex} \equiv N_{\rm thermal}/N_{\rm tot}$ are about $100$ nK and $15 \sim 20\%$, respectively. We stress that these quantities, together with the average values of $T$ and $P_{\rm ex}$ for different lattice depth, are all directly measured from experiments with no fitting parameters.

We consider separately the tilt of OL, the fraction of excited atoms, and the width variation of interference peaks induced by defocusing effect, and compare the corresponding numerical results with experimental outcome of $P_3$, $P_4$, and $P_5$, respectively. In Figs.~\ref{fig:ol_pca_img}(d2), \ref{fig:ol_pca_img}(e2) and \ref{fig:ol_pca_img}(f2), blue lines are experimental results and orange lines are simulation results. The details of our simulation results are as follows.

\paragraph{Potential energy gradient in OL}
As shown in Fig.~\ref{fig:ol_pca_img}(d), a translational shift along the \textit{x}-direction is mainly manifested by $P_3$. In the absence of OL, the eigenvalue of $P_3$ is of the same order in magnitude as that of $P_2$, which characterizes position fluctuations along the \textit{z}-direction (see Fig.~\ref{fig:bec_pca_contrast_new_img}). In an OL, however, $P_3$ becomes more significant for most experimental realizations. This strongly indicates that there must be some effects related to the OL contribute to $P_3$, in addition to the spatial shifts of the magnetic trap, the BEC and the camera.

We attribute this fluctuation to a potential energy gradient between different lattice sites. One of the major gradient sources comes from the gravity field as the OL cannot be perfectly horizontal, which introduces a phase gradient 
\begin{eqnarray}
\Delta\phi = mg\sin\theta\cdot\lambda{t}_{\rm hold}/2\hbar
\end{eqnarray} 
between sites separated by $\lambda/2$, where $m$, $g$, $t_{\rm hold}$, and $\theta$ are the atom mass, the gravity acceleration constant, the holding time of OL, and the angle of tilting from the horizontal direction, respectively. As a result, the TOF pattern is shifted along the direction of OL. In our numerical simulation, we consider a random tilting with  an angle of no more than $\pm1.8\times 10^{-5}$ rad. We find a good agreement between the simulation and the experimental result, as shown in Fig.~\ref{fig:ol_pca_img}(d2). This observation suggests that the PCA method can reveal very small potential gradient in OL, which may have application in detecting microgravity. Although the sensitivity reported here is about two orders of magnitude smaller than the uncertainty of $10^{-7}$ reached by measuring the 5th harmonic of Bloch oscillation of $^{88}$Sr atoms in tilted optical lattices~\cite{blochgravity}, our scheme can be easily implemented with $^{87}$Rb atoms in a simpler experimental setup with conventional TOF techniques.

\paragraph{Fraction of excited atoms}
In our experiment, the fraction of normal state atoms can hardly be a constant because of the imperfection of our preparation and loading processes~\cite{McKagan}. It is reflected by ${P}_5$ in Fig.\ref{fig:ol_pca_img}(f1). We emphasize that to obtain a quantitative agreement with the experimental observation of $P_5$, in the numerical simulation we consider fluctuation of excited fraction, under a constraint of total particle number conservation. This requires a normalization of the solution of GPE after a Gaussian fluctuation is added. 

\paragraph{Width variation caused by defocusing effect}
The GPE is a nonlinear equation with an interaction potential term $N{U}_{0}{\left|\psi\left(\mathbf{r},t\right)\right|}^{2}\psi\left(\mathbf{r},t\right)$. This repulsive interaction between atoms tends to broaden the atomic distribution in both spatial and momentum space, resulting a defocusing effect with wider peak widths in TOF images. Thus, the variation of atom number $N$ can induce fluctuations to TOF signals, which can not be fully eliminated by a normalization of the BEC wave function.  Another factor one needs to take into account is that while reducing three-dimensional GPE to 1D GPE, we have to assume a distribution along the $y$- and $z$-directions to reduce ${\left|\psi\left(\mathbf{r},t\right)\right|}^{2}$ to ${\left|\psi\left(x,t\right)\right|}^{2}$. This means the density distributions along the $y$- and $z$-directions still have an influence on peak width along the $x$-direction. Fluctuation in these transversal directions as shown in Fig.~\ref{fig:bec_pca_img}(d) for the case without OL is hence another source of noise.

\subsection{\label{subsec:depth_trend}Variations with OL depth}

We now turn to the PCA results for OLs of different depths. To quantify the trend of variation of PCs,
we define a quantity
\begin{eqnarray}
\gamma_{P_i}=\frac{E_{P_i}}{\sum_{j=1}^{5}E_{P_j}},
\end{eqnarray} 
where $E_{P_i}$ is the eigen value of $P_i$. In the following discussion, we combine $P_2$ and $P_3$ together and study $\gamma_{P_2 + P_3} \equiv \gamma_{P_2} + \gamma_{P_3} $ because their underlying physical origins are the same.

\begin{figure}[tbp]
\centering\includegraphics[width=11cm]{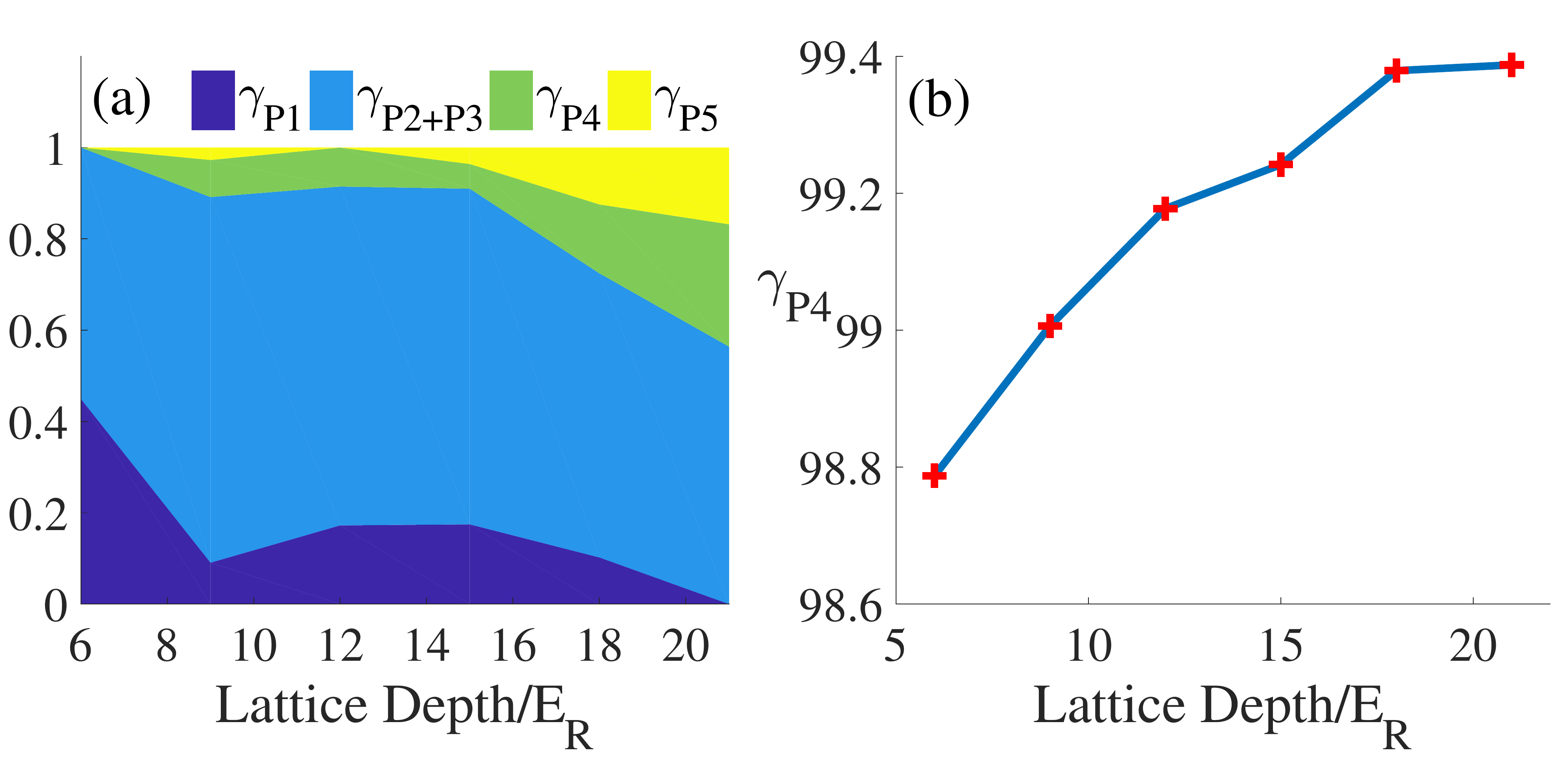}
\caption{(a) Trends of PCs as lattice depth increases in experiments. We add up portions of $P_2$ and $P_3$ because they have the same physical origin. (b) The trend of $\gamma_{P_4}$ from numerical simulation. The red crossings are results of configurations we chose to simulate and the blue line is a smoothed line that connects our results. In our simulation, we use the values extracted from TOF images for the mean and fluctuation of thermal atom fraction and temperature.}
\label{fig:simu_illus}
\end{figure}

From Fig.~\ref{fig:simu_illus}(a), we notice that both $\gamma_{P_1}$ and $\gamma_{P_2+P_3}$ decrease 
with increasing lattice depth, while $\gamma_{P_4}$ and $\gamma_{P_5}$ exhibit an opposite dependence. 
These trends are qualitatively consistent with our attribution of physical origins of noises, considering the fact that the fluctuation of excited fraction ($P_4$) and the interaction effect ($P_5$) become more severe
as OL gets deeper. In fact, we numerically simulate peak width variation under density distribution fluctuation 
at different lattice depths with all other conditions fixed. As shown in Fig.~\ref{fig:simu_illus}(b), $\gamma_{P_4}$ indeed grows when approaching the quantum phase transition with increasing OL depth, which agrees with the experiment result qualitatively. 
In fact, one would naturally expect that fluctuation of interference between different lattice sites will be significantly enhanced near the phase transition point.

\section{\label{pca_discussion}Summary and final remarks}

Before concluding, we emphasize that the eigenvectors given by PCA are uncorrelated but not necessarily independent with respect to physical noises. In other words, one single PC can reflect a combination of more than one physical origins, although in most cases only one source is dominating and can be clearly distinguished. In this paper, we demonstrate that PCA can be readily implemented to obtain accurate physical interpretations for noises in complex many-body systems. As a comparison, independent component analysis (ICA), a method based on PCA, can reach optimized and parameterized independence between basis vectors~\cite{Segal}. But it requires a clear understanding about the sources of noise and their effects on the system to properly set criteria of independence. An inappropriate parameter setting could lead to severe wrongful analysis. This kind of misleading is absent in PCA as it is a standard, non-parametric method, requiring no prior knowledge of the system.

In summary, we introduce a method of PCA to analyze noise in TOF images of a BEC in a 1D OL. 
By investigating the corresponding feature images, we identify the physical origins associated to a few PCs of leading contribution. This understanding is then confirmed by a numerical simulation of a GPE with external sources of fluctuations. In particular, we can extract not only classical fluctuations such as a small tilt angle of the optical lattice laser, but also quantum fluctuations such as the fraction of non-condensed excited particles. Both factors are very weak effects that can not be extracted by conventional investigation of interference patterns. Based on the knowledge of the physical origins of leading PCs, we also design a preprocessing method to significantly reduce or even eliminate fluctuations of atom number and spatial position. The PCA method could find plausible applications in the future, including interferometers with higher precision, measurement of microgravity, and high precision level meters.

\section*{Funding}

National Key Research and Development Program of China (Grant No. 2016YFA0301501); National Science Foundation of China (Grant Nos. 11334001, 61475007, 61727819, 91736208, 11522436, 11774425); Research Funds of Renmin University of China (Grant No. 16XNLQ03).

\bibliography{pca_reference}






\end{document}